\documentclass[12pt,twoside]{article}
\usepackage{amsmath}
\usepackage{amssymb}
\usepackage{array}
\usepackage{graphicx}
\usepackage{geometry}
\usepackage{graphics}
\usepackage{enumerate}
\usepackage[utf8]{inputenc}

\begin{document}
\vskip7cm\noindent

\setcounter{page}{1}
\font\tinyfont=cmr8
\font\headd=cmr8
\pagestyle{myheadings}
\font\bigfont=cmssbx10 at 18pt
\font\tinyfont=cmr8
\font\bigfont=cmssbx10 at 18pt
\begin{center}{\bf Pathway to Fractional Integrals, Fractional Differential Equations and the Role of H-function}\end{center}

\vskip1cm\begin{center}A.M. MATHAI\\
Department of Mathematics and Statistics\\
McGill University, Montreal, Canada, H3A 2K6\\
directorcms458@gmail.com\\
\&\\
H.J. HAUBOLD\\
Office for Outer Space Affairs, United Nations\\
Vienna International Centre, Box 500, A-1400 Vienna, Austria\\
hans.haubold@gmail.com\\
\end{center}

\vskip.5cm\noindent{\bf Abstract}

\vskip.3cm Pathway model for the real scalar variable case is re-explored and its connections to fractional integrals, solutions of fractional differential equations, Tsallis statistics and superstatistics in statistical mechanics, reaction-rate probability integral, Kr\"atzel transform, pathway transform etc are explored. It is shown that the common thread in these connections is the H-function representations. The pathway parameter is shown to be connected to the fractional order in fractional integrals and fractional differential equations.

\vskip.3cm\noindent{\bf Keywords:} Pathway model, fractional integral, fractional differential equations, reaction-rate probability integral, Kr\"atzel integral, pathway transform.

\vskip.3cm\noindent{\bf MSC2020 Numbers:} 26A33, 26B15, 33C99, 33E12, 49K05, 62E15, 94A15

\vskip.5cm\noindent{\bf 1.\hskip.3cm Introduction}

\vskip.3cm The pathway model was introduced for the real rectangular matrix-variate case in Mathai (2005). Later it was extended to cover the complex rectangular matrix-variate case in Mathai and Provost (2006). The real scalar version of the model is the following:
$$f_1(x)=c_1|x|^{\gamma}[1-a(1-q)|x|^{\delta}]^{\frac{\eta}{1-q}}\eqno(1.1)
$$for $a>0,q<1,\gamma>-1,\delta>0, \eta>0, 1-a(1-q)|x|^{\delta}>0$, and zero elsewhere,  where $c_1$ can act as the normalizing constant if we wish to create a statistical density out of $f_1(x)$. The model in (1.1) can be looked upon as a generalized and extended type-1 beta model for $q<1$. The functional part in the basic type-1 beta model is $x^{\alpha-1}(1-x)^{\beta-1},0\le x\le 1,\alpha>0, \beta>0$ and zero elsewhere and hence (1.1) is the generalized and extended form of this basic model. Note that (1.1) is a symmetric model for $x<0$ and $x>0$. If an asymmetric model is required then different weights can be used for $x<0$ and $x>0$ situations. These differing weights can be introduced either by multiplying with constants or through one of the parameters.
\vskip.2cm For $q>1$, write $1-q=-(q-1), q>1$, then (1.1) switches into the following model:
$$f_2(x)=c_2|x|^{\gamma}[1+a(q-1)|x|^{\delta}]^{-\frac{\eta}{q-1}}\eqno(1.2)
$$for $-\infty<x<\infty,a>0, q>1, \delta>0, \gamma>-1, \eta>0$. The model in  (1.2) can be looked upon as a generalized  and extended type-2 beta family of functions. The functional part of the basic type-2 beta model is $x^{\alpha-1}(1+x)^{-(\alpha+\beta)},0\le x<\infty,\alpha>0,\beta>0$. Thus, (1.2) is the generalized and extended version of this basic type-2 beta function. When $q\to 1_{-}$ in (1.1) and $q\to 1_{+}$ in (1.2) the models in (1.1) and (1.2) go to the generalized and extended gamma family of functions as follows:
$$f_3(x)=c_3|x|^{\gamma}{\rm e}^{-a\eta |x|^{\delta}}\eqno(1.3)
$$for $a>0, \eta>0, \delta>0, \gamma>-1, -\infty<x<\infty$. Thus, the basic pathway model is (1.1), and (1.2) and (1.3) are available from (1.1). For $q<1$ the family of functions is the generalized and extended type-1 beta family of functions. When $q$ goes to $1$, then one goes into the generalized and extended gamma family of functions. When $q$ moves to $q>1$ then we go into the generalized and extended type-2 beta family of functions. The pathway idea in model building situation is to capture the stable as well as the unstable neighborhoods by the same model. If the gamma family, such as a Maxwell-Boltzmann density or a Raleigh density, is the stable situation in a physical problem then the paths leading to this stable situation through the generalized type-1 beta model and generalized type-2 beta model and the transitional stages are captured by the pathway parameter $q$. If (1.1) to (1.3) are to be treated as statistical densities then the following are the normalizing constants:
\begin{align*}
c_1&=\delta[a(1-q)]^{\frac{\gamma+1}{\delta}}\frac{\Gamma(\frac{\eta}{1-q}+1+\frac{\gamma+1}{\delta})}
{2\Gamma(\frac{\gamma+1}{\delta})\Gamma(\frac{\eta}{1-q}+1)}\tag{1.4}\\
&\mbox{for  }\gamma>-1,a>0,q<1,\eta>0,\delta>0\\
c_2&=\delta[a(q-1)]^{\frac{\gamma+1}{\delta}}\frac{\Gamma(\frac{\eta}{q-1})}
{2\Gamma(\frac{\gamma+1}{\delta})\Gamma(\frac{\eta}{q-1}-\frac{\gamma+1}{\delta})}\tag{1.5}\\
&\mbox{for  }\eta>0,\delta>0,a>0,q>1,\gamma>-1,\frac{\eta}{q-1}-\frac{\gamma+1}{\delta}>0\\
c_3&=\frac{\delta (a\eta)^{\frac{\gamma+1}{\delta}}}{2\Gamma(\frac{\gamma+1}{\delta})},a>0,\eta>0,\delta>0.\tag{1.6}
\end{align*}
\vskip.2cm Note that (1.1) for $\gamma=0,a=1,\delta=1,\eta=1, q<1,q>1,q\to 1$ is Tsallis statistics in non-extensive statistical mechanics (Tsallis, 2009). Equation (1.2) for $q>1,q\to 1,a=1,\delta=1,\eta=1$ is superstatisitcs in statistical mechanics (Beck and Cohen, 2003).\vskip.2cm
If location and scale parameters are to be incorporated into the model then in (1.1) to (1.3), replace $x$ by $\frac{x-\mu}{\sigma}$ for some parameter $\mu$ and $\sigma>0$. Note that (1.3) for $\gamma=0,\delta=2$ is the normal or Gaussian density. For $x>0$, (1.3) produces the generalized gamma density, Weibull density, gamma density, chisquare density, exponential density. Rayleigh density, Maxwell-Boltzmann density etc.\vskip.2cm
Exponentiation in (1.2), that is put $x={\rm e}^{-cy},c>0$, produces the generalized logistic density of Mathai and Provost (2006a), logistic density and other special cases of the generalized logistic density. (1.2) can produce Cauchy density, Student-t density, F-density etc. As a limiting form one can obtain Fermi-Dirac density from (1.2) and Bose-Einstein density from (1.1), after exponentiation.

\vskip.3cm\noindent{\bf 2.\hskip.3cm Optimization of Entropy}

\vskip.3cm In (1.1) to (1.3) we have the distributional pathway or pathway through densities. We can also provide an entropic pathway by deriving (1.1) through optimization of an entropy measure. Consider Mathai's entropy, for an earlier version see Mathai and Haubold (2007),
$$M_q(f)=\frac{\int_X[f(X)]^{\frac{1-q+\eta}{\eta}}{\rm d}X-1}{q-1},q\ne 1,q<1+\eta\eqno(2.1)
$$where $f(X)$ is a statistical density where $X$ could be real or complex scalar or matrix variable, and $\int_X$ denotes the integral over the support of $f$. For $\eta=1$, one can look upon (2.1) as an expected value of $f^{1-q}$ which then corresponds to Kerridge's measure of inaccuracy, see Mathai and Rathie (1975). In general, (2.1) is the expected value of $f^{\frac{1-q}{\eta}}$. In the following discussion we consider the real scalar variable case first. Let us optimize (2.1) subject to the following moment-type restrictions for real scalar variables:
$$\int_xx^{\gamma\frac{(1-q)}{\eta}}f(x){\rm d}x=\mbox{ fixed }\eqno(2.2a)$$
and
$$\int_xx^{\gamma\frac{(1-q)}{\eta}+\delta}f(x){\rm d}x=\mbox{ fixed }\eqno(2.2b)
$$for $\gamma>-1,q<1,\delta>0$, and it is assumed that (2.2a) and (2.2b) exist. Note that for $\gamma=0$, (2.2a) is equivalent to the statement that the total integral is unity and (2.2b) for $\delta=1,\gamma=0$ states that the first moment is given. If we use calculus of variation for the optimization of (2.1) then the Euler equation is the following:
\begin{align*}
\frac{\partial}{\partial f}[f^{\frac{1-q+\eta}{\eta}}&-\lambda_1x^{\gamma\frac{(1-q)}{\eta}}f
+\lambda_2x^{\gamma\frac{(1-q)}{\eta}+\delta}f]=0\tag{2.3}\\
&\mbox{where $\lambda_1$ and $\lambda_2$ are Lagrangian multipliers}\Rightarrow\\
f^{\frac{(1-q)}{\eta}}&=\mu_1x^{\gamma\frac{(1-q)}{\eta}}[1-\mu_2x^{\delta}]\\
&\mbox{for some constants $\mu_1$ and $\mu_2$, which then gives}\\
f_1&=\gamma_1x^{\gamma}[1-\mu_2 x^{\delta}]^{\frac{\eta}{1-q}}\tag{2.4}
\end{align*}for some $\gamma_1$ and $\mu_2$. Take $\mu_2=a(1-q)$ and $\gamma_1$ as the normalizing constant to obtain the model (1.1). The entropy in (2.1) is a modified version of Havrda-Charvat $\alpha$-generalized entropy, see Mathai and Rathie (1975). Another modified form of Havrda-Charvat entropy is Tsallis entropy (Tsallis, 2009), namely
$$T(f)=\frac{\int_x[f(x)]^q{\rm d}x-1}{1-q},q\ne 1\eqno(2.5)
$$where $f(x)$ is the density of the real scalar random variable $x$. Optimization of (2.5) under the restrictions (2.2a) and (2.2b) for $\gamma=0,\delta=1$ in an associated escort density will produce Tsallis statistics. Direct optimization of (2.5) will produce $q$-exponential function. But the pathway model is available directly from (2.1) under the restrictions (2.2a) and (2.2b) from where Tsallis statistics and superstatistics are available as special cases. Superstatistics considerations can be explained in terms of  statistical language as the construction of an unconditional density when the conditional density and marginal density belong to generalized gamma families of functions. Such a procedure can produce only the cases $q>1$ and $q\to 1$, not the case $q<1$. But optimization of (2.5) through an escort density can produce special cases of the pathway model for all the cases $q<1,q>1,q\to 1$. This is the advantage of Tsallis statistics over superstatistics.
\vskip.2cm In (2.1) if $X$ is a $p\times 1$ vector random variable and  if (2.2a) and (2.2b) are replaced by the following conditions
$$\int_X[(X-\mu)'V^{-1}(X-\mu)]^{\gamma\frac{(q-1)}{\eta}}f(X){\rm d}X=\mbox{ fixed }\eqno(2.6a)$$and
$$\int_X[(X-\mu)'V^{-1}(X-\mu)]^{\gamma\frac{(q-1)}{\eta}+\delta}f(X){\rm d}X=\mbox{ fixed }\eqno(2.6b)
$$where $V$ is $p\times p$ real symmetric and positive definite constant matrix, $\mu$ is a $p\times 1$ constant vector, a prime denotes transpose, $\gamma>0,q>1,\eta>0$, then from (2.3) and (2.4) we have the following density:
$$f_4(X)=c_4[(X-\mu)]V^{-1}(X-\mu)]^{\gamma}[1+a(q-1)\{(X-\mu)'V^{-1}(X-\mu)\}^{\delta}]^{-\frac{\eta}{q-1}},q>1.\eqno(2.7)
$$Then, when $q\to 1$, $f_4(X)$ goes to $f_5(X)$ given by
$$f_5(X)=c_5[(X-\mu)'V^{-1}(X-\mu)]^{\gamma}{\rm e}^{-a\eta[(X-\mu)'V^{-1}(X-\mu)]^{\delta}}.\eqno(2.8)
$$Note that (2.7) and (2.8) are also associated with type-2 and gamma distributed random points in Euclidean $n$-space, $p\le n$, see Mathai (1999) for details. Also, (2.8) for $\gamma=0,\delta=1$ is the $p$-variate Gaussian density with mean value vector $\mu$ and covariance matrix $V$.
\vskip.2cm In (2.1) if $X$ is a $p\times q$, $q\ge p$ rectangular matrix-variate random variable and if the conditions in (2.2a) and (2.2b) are replaced by the following:
$$\int_X[{\rm tr}(AXBX')]^{\gamma\frac{(q-1)}{\eta}}f(X){\rm d}X=\mbox{ fixed }\eqno(2.9a)
$$and
$$
\int_X[{\rm tr}(AXBX')]^{\gamma\frac{(q-1)}{\eta}+\delta}f(X){\rm d}X=\mbox{ fixed }\eqno(2.9b)
$$where $A$ is $p\times p$ and $B$ is $q\times q$ constant real positive definite matrices, $q>1,\eta>0,\delta>0$ then (2.3) and (2.4) will give the density
$$f_6(X)=c_6[{\rm tr}(AXBX')]^{\gamma}[1+a(q-1)\{{\rm tr}(AXBX')\}^{\delta}]^{-\frac{\eta}{q-1}},q>1\eqno(2.10)
$$which when $q\to 1$ gives
$$f_7(X)=c_7[{\rm tr}(AXBX')]^{\gamma}{\rm e}^{-a\eta[{\rm tr}(AXBX')]^{\delta}}\eqno(2.11)
$$for $a>0,\eta>0,\delta>0,\gamma>-1$. Note that (2.11) for $\gamma=0,\delta=1$ is the real matrix-variate Gaussian density. By replacing $X$ by $X-M$, where $M$ is a $p\times q$ constant matrix, one can also incorporate a location parameter matrix in the models in (2.10) and (2.11).

\vskip.5cm\noindent{\bf 3.\hskip.3cm Connection of Pathway Model to Fractional Integrals}

\vskip.3cm From the geometrical interpretation of the fractional integral given in Mathai (2014) it is evident that a type-1 beta form must be present in the definition of a fractional integral if we wish to encompass all the different definitions of fractional integrals in current use. Hence a definition, covering all fractional integrals in current use, introduced in Mathai (2013, 2014, 2015), is as Mellin convolutions of products and ratios where the functions involved in the Mellin convolutions are of the following form for the real scalar variable case $x$:

$$f_1(x_1)=\phi_1(x_1)[1-a(1-q)x_1^{\delta}]^{\frac{1}{1-q}}~~\&~~ f_2(x_2)=\phi_2(x_2)f(x_2)\eqno(3.1)
$$where $\phi_1$ and $\phi_2$ are prefixed functions and $f$ is an arbitrary function, $a>0,\delta>0,q<1$. If statistical densities are needed then multiply $f_1$ and $f_2$ by appropriate normalizing constants. In this case $\phi_1,\phi_2,f$ are to be restricted to be positive functions and $1-a(1-q)x^{\delta}>0$. Mellin convolution of products will correspond to fractional integrals of the second kind or the right-sided integrals and Mellin convolution of ratios will correspond to fractional integrals of the first kind or left-sided integrals.
\vskip.5cm\noindent{\bf 3.1.\hskip.3cm Fractional integrals of the second kind}

\vskip.3cm Let $u=x_1x_2, v=x_2$ or $x_2=v, x_1=\frac{u}{v}$ and the Jacobian is $\frac{1}{v}$. Then the Mellin convolution of the product, denoted by $g_2(u)$, is the following, taking the pathway model of (3.1) as $f_1(x_1)$ with $\phi_1(x_1)=x_1^{\gamma}$, $\phi_2(x_2)=1$.
\begin{align*}
g_2(u)&=\int_v\frac{1}{v}(\frac{u}{v})^{\gamma}[1-a(1-q)(\frac{u}{v})^{\delta}]^{\frac{1}{1-q}}f(v){\rm d}v,q<1\tag{3.2}\\
&=u^{\gamma}\int_v v^{-\gamma-1}[1-a(1-q)(\frac{u^{\delta}}{v^{\delta}})]^{\frac{1}{1-q}}f(v){\rm d}v.\end{align*}
Let us consider the Mellin transform of $g_2(u)$,  with Mellin parameter $s$. Then

$$M_{g_2}(s)=\int_{u=0}^{\infty}u^{\gamma+s-1}\int_{v>[a(1-q)]^{\frac{1}{\delta}}u}
[1-a(1-q)\frac{u^{\delta}}{v^{\delta}}]^{\frac{1}{1-q}}f(v){\rm d}v.\eqno(3.3)
$$Let us integrate out $u$ first. Then $0\le u\le \frac{v}{[a(1-q)]^{\frac{1}{\delta}}}$.  Integral over $u$ gives the following:
\begin{align*}
\int_{u=0}^{\frac{v}{[a(1-q)]^{\frac{1}{\delta}}}}u^{\gamma+s-1}&
[1-a(1-q)\frac{u^{\delta}}{v^{\delta}}]^{\frac{1}{1-q}}{\rm d}u\\
&=\frac{1}{\delta}\frac{v^{\gamma+s}}{[a(1-q)]^{\frac{\gamma+s}{\delta}}}
\int_0^1t^{\frac{\gamma+s}{\delta}-1}[1-t]^{\frac{1}{1-q}}{\rm d}t\\
&=\frac{1}{\delta}\frac{\Gamma(\frac{1}{1-q}+1)}{[a(1-q)]^{\frac{\gamma+s}{\delta}}}
\frac{\Gamma(\frac{\gamma}{\delta}+\frac{s}{\delta})}{\Gamma(\frac{1}{1-q}
+\frac{\gamma}{\delta}+\frac{s}{\delta})}
\end{align*}for $\Re(\gamma+s)>0,\delta>0,a>0,q<1$ where $\Re(\cdot)$ means the real part of $(\cdot)$. Now taking the integral over $v$ we have the Mellin transform of the arbitrary function $f(v)$, denoted by $f^{*}(s)$. Therefore

$$M_{g_2}(s)=\frac{\Gamma(\frac{1}{1-q}+1)}{\delta[a(1-q)]^{\frac{\gamma+s}{\delta}}}
\frac{\Gamma(\frac{\gamma+s}{\delta})}{\Gamma(\frac{1}{1-q}+1+\frac{\gamma+s}{\delta})}f^{*}(s)\eqno(3.4a)
$$for $\Re(\gamma+s)>0,a>0,\delta>0, q<1$;
$$M_{g_2}(s)=\frac{\Gamma(\frac{1}{1-q}+1)}{\delta(1-q)^{\frac{s}{\delta}}}
\frac{\Gamma(\frac{s}{\delta})}{\Gamma(\frac{1}{1-q}+1+\frac{s}{\delta})}f^{*}(s)\mbox{ for } a=1,\gamma=0;\eqno(3.4b)
$$
$$M_{g_2}(s)=\frac{\Gamma(\frac{\gamma+s}{\delta})}{\delta a^{\frac{\gamma+s}{\delta}}}f^{*}(s)\mbox{ for }\delta>0,a>0,q\to 1.\eqno(3.4c)
$$Now, compare (3.4a) with Theorem 3.4.2 of the Mellin transform of Erd\'{e}lyi-Kober fractional integral of the second kind for $\delta=1$, see Mathai (2014a). It is the same structure and $\frac{1}{1-q}+1$ corresponds to the order of the fractional integral $\alpha$. Compare (3.4b) with the Mellin transform of Weyl fractional integral of the second kind for $\delta=1$ (Mathai, 2014a). Again, the fractional order $\alpha$ corresponds to $\frac{1}{1-q}+1$.

\vskip.5cm\noindent{\bf 3.2.\hskip.3cm Fractional integral of the first kind or left-sided integral}

\vskip.3cm Let $u=\frac{x_2}{x_1}, v=x_2$ or $x_2=v,  x_1=\frac{v}{u}$. Jacobian is $-\frac{v}{u^2}$. Let $\phi_1(x_1)=x_1^{\gamma-1}, \phi_2(x_2)=1$. Let the Mellin convolution of the ratio be denoted by $g_1(u)$. Then
\begin{align*}
g_1(u)&=\int_v\frac{v}{u^2}(\frac{v}{u})^{\gamma-1}
[1-a(1-q)(\frac{v}{u})^{\delta}]^{\frac{1}{1-q}}f(v){\rm d}v\\
&=u^{-\gamma-1-\frac{\delta}{1-q}}\int_v v^{\gamma}[u^{\delta}-a(1-q)v^{\delta}]^{\frac{1}{1-q}}f(v){\rm d}v.\tag{3.5}\end{align*}
Mellin transform of $g_1(u)$, with Mellin parameter $s$, is the following:

$$M_{g_1}(s)=\int_{u=0}^{\infty}u^{-\gamma-1+s-1-\frac{\delta}{1-q}}
\int_v v^{\gamma}[u^{\delta}-a(1-q)v^{\delta}]^{\frac{1}{1-q}}f(v){\rm d}v.
$$Integrating out $u$ first, we have the integral over $u$ as follows:
\begin{align*}
\int_{u=[a(1-q)]^{\frac{1}{\delta}}v}^{\infty}&[u^{\delta}-a(1-q)v^{\delta}]^{\frac{1}{1-q}}
u^{-\gamma-1+s-1-\frac{\delta}{1-q}}{\rm d}u\\
&=\frac{1}{\delta}\int_{z=0}^{\infty}z^{\frac{1}{1-q}}[z+a(1-q)v^{\delta}]^{-\frac{1}{\delta}(\gamma+1+\frac{\delta}{1-q}-s)-1}{\rm d}z.\end{align*}
This is evaluated with the help of a type-2 beta integral. Then integration over $v$ gives the following final result:
$$M_{g_1}(s)=\frac{\Gamma(\frac{1}{1-q}+1)}{[a(1-q)]^{\frac{\gamma+1-s}{\delta}}}
\frac{\Gamma(\frac{\gamma+1-s}{\delta})}
{\Gamma(\frac{1}{1-q}+1+\frac{\gamma+1-s}{\delta})}f^{*}(s),\Re(\gamma+1-s)>0.\eqno(3.6)
$$The fractional order $\alpha$ corresponds to $\frac{1}{1-q}+1$, as seen before. The pathway parameter for $q<1$ is such that
$$\frac{1}{1-q}+1=\alpha,q<1
$$where $\alpha$ is the fractional order in the fractional integrals of the first and second kinds.

\vskip.5cm\noindent{\bf 4.\hskip.3cm Mellin Convolutions of Products and Ratios for Other Functions}

\vskip.3cm When $f_1$ is connected to pathway model for $q<1$ and $f_2$ an arbitrary function then the Mellin convolutions of products and ratios are seen to be connected to fractional integrals of the second and first kind, respectively. Let $f_1$ and $f_2$ be generalized gamma functions which are actually the pathway model for $q\to 1$. Take (1.1) for $q\to 1_{-}$ and (1.2) for $q\to 1_{+}$. Let us see what happens to the Mellin convolutions of products and ratios. Let $u=x_1x_2, v=x_1, x_1=\frac{u}{v}$. Jacobian is $\frac{1}{v}$. Let

$$f_1(x_1)=C_1x^{\gamma_1-1}{\rm e}^{-a_1x_1^{\delta_1}}~~\&~~ f_2(x_2)=C_2x_2^{\gamma_2-1}{\rm e}^{-a_2x_2^{\delta_2}}\eqno(4.1)
$$for $\gamma_j>0,a_j>0,\delta_j>0,j=1,2$, where $C_1$ and $C_2$ can be normalizing constants if $f_j(x_j),j=1,2$ are to be treated as statistical densities. Let the Mellin convolution of a product be again denoted by $g_2(u)$. Then
\begin{align*}
g_2(u)&=C_1C_2\int_v\frac{1}{v}(\frac{u}{v})^{\gamma_1-1}v^{\gamma_2-1}{\rm e}^{-a_1(\frac{u}{v})^{\delta_1}-a_2v^{\delta_2}}\\
&=C_1C_2u^{\gamma_1-1}\int_{v=0}^{\infty}v^{\gamma_2-\gamma_1-1}{\rm e}^{-\frac{a_1u^{\delta_1}}{v^{\delta_1}}-a_2v^{\delta_2}}{\rm d}v\tag{4.2}\end{align*}
For $\delta_1=1,\delta_2=1$, (4.2) gives the basic Kr\"atzel integral, and associated with it we have the Kr\"atzel transform (Kr\"atzel, 1979). See Mathai (2012) for some properties of generalized Kr\"atzel integral and its connection to statistical distribution theory and connection to reaction rate probability integral in nuclear reaction rate theory (Mathai and Haubold, 1988)]. For $\delta_2=1,\delta_1=\frac{1}{2}$, (4.2) gives the reaction rate probability integral. For $\delta_1=1,\delta_2=1$ and $\gamma_2-\gamma_1=-\frac{1}{2}$ the integrand in (4.2) gives the inverse Gaussian density in stochastic processes. The structure in (4.2) is also that of the  unconditional density in a Bayesian set up with the conditional density belonging to the generalized gamma family of the type $b_1x_1^{\gamma_1}{\rm e}^{-a{x_1^{\delta_1}}/{x_2^{\delta_1}}}$ and the marginal density also belonging to generalized gamma family of the form $x_2^{\gamma_2}{\rm e}^{-bx_2^{\delta_2}}$. Then the unconditional density is of the form of the integral in (4.2). Observe that the generalized gamma functions that we considered in (4.2) are nothing but the limiting forms of the pathway models in (1.1) and (1.2) for $q\to 1$ or the model in (1.3).

\vskip.5cm\noindent{\bf 5.\hskip.3cm Connection of Pathway Parameter to Fractional Indices in Fractional Differential Equations}

\vskip.3cm For illustrative purposes let us consider the fractional space-time diffusion equation, see Haubold et al. (2011a).

$${_0D_t^{\beta}}N(x,t)=\eta~{_xD_\theta^{\alpha}}N(x,t)\eqno(5.1)
$$with the initial condition ${_0D_x}^{\beta-1}N(x,0)=\sigma(x), 0\le\beta\le 1, \lim_{x\to\pm\infty}N(x,t)=0$, $\eta$ is the diffusion constant. $\eta,t>0, x\in R, \alpha,\theta,\beta$ are real parameters, $0<\alpha\le 2,|\theta|\le \min\{\alpha,2-\alpha\}$. The solution of (5.1) for the case $\alpha=\beta$ is available from Haubold et al. (2011a) in the following form for $\rho=\frac{\alpha-\theta}{2\alpha}$:

\begin{align*}
N(x,t)&=\frac{t^{\beta-1}}{\alpha|x|}H_{3,3}^{2,1}\left[\frac{|x|}{t\eta^{\frac{1}{\alpha}}}\bigg\vert_{(1,\frac{1}{\alpha}),(1,1),(1,\rho)}^{(1,\frac{1}{\alpha}),(\alpha,1),(1,\rho)}\right]\tag{5.2}\\
&=\frac{t^{\alpha-1}}{\alpha|x|}\frac{1}{2\pi i}\int_L\frac{\Gamma(1+\frac{s}{\alpha})\Gamma(1+s)\Gamma(-\frac{s}{\alpha})}
{\Gamma(1-\rho s)\Gamma(\alpha+s)\Gamma(1+\rho s)}\left[\frac{|x|}{t\eta^{\frac{1}{\alpha}}}\right]^{-s}{\rm d}s\end{align*}, where $L$ is an appropriate contour and $i=\sqrt{-1}$. Let
\begin{align*}
g(x,t)&=\frac{\alpha|x|}{t^{\beta-1}}N(x,t)\\
&=\frac{1}{2\pi i}\int_L\frac{\Gamma(1+\frac{s}{\alpha})\Gamma(-\frac{s}{\alpha})}{\Gamma(-\rho s)\Gamma(1-\rho+\rho s)}\frac{\Gamma(1+s)}{\Gamma(\alpha+s)}\left[\frac{|x|}{t\eta^{\frac{1}{\alpha}}}\right]^{-s}{\rm d}s.\tag{5.3}
\end{align*}
Therefore, the Mellin transform of $g(x,t)$, with Mellin parameter $s$, and argument $\frac{|x|}{t\eta^{\frac{1}{\alpha}}}$ is given by the following:

$$M_{g}(s)=\frac{\Gamma(1+s)}{\Gamma(\alpha+s)}f^{*}(s), ~~f^{*}(s)=\frac{\Gamma(1+\frac{s}{\alpha})\Gamma(-\frac{s}{\alpha})}{\Gamma(-\rho s)\Gamma(1+\rho s)}.\eqno(5.4)
$$Now compare (5.4) with (3.4b) for $\delta=1$. The pathway integrals of the second kind for $q<1$ is of the form $c\frac{\Gamma(1+s)}{\Gamma(\alpha+s)}f^{*}(s)$ where $c$ is a constant and $\alpha=\frac{1}{1-q}+1$  is the order of the fractional integral of the second kind and $\alpha$ is also order of the fractional differential equation. But (5.4) is the Mellin transform of the solution of fractional space-time diffusion equation for the case $\alpha=\beta$ where $\alpha$ is the fractional index in the fractional differential equation. This is the same $\alpha$ appearing in (5.4) which is also equal to $\frac{1}{1-q}+1=\alpha$. This is the same $\alpha$ entering as the coefficient of $s$ in the H-function representation, the coefficients being $\frac{1}{\alpha}$, and $\rho$ which is a function of $\alpha$. Note also that in (5.4), $f^{*}(s)$ has an interesting structure. Both the numerator and denominator are of the form $\Gamma(z)\Gamma(1-z)$ for different $z$. Hence the gamma product can be written in terms of $\sin\pi z$.

\vskip.5cm\noindent{\bf 6.\hskip.3cm The H-function Thread}

\vskip.3cm What we have seen is that the pathway model, fractional integrals of the first and second kinds, and fractional differential equation are all connected through the pathway model  and the basic representation is in terms of the H-function. Solutions of simple fractional differential equations are available in terms of Mittag-Leffler functions. Mittag-Leffler functions are special cases of the H-function. For an overview of Mittag-Leffler function and its properties, see Haubold et al. (2011). Mittag-Leffler functions, as solutions of fractional differential equations, may be seen from Gorenflo et al. (2000), Carpinteri and Mainardi (1997), Kiryakova (1994). For example, a three-parameter Mittag-Leffler function has the H-function representation:

\begin{align*}
E_{\alpha,\beta}^{\gamma}(z)&=\frac{1}{2\pi i}\int_{c-i\infty}^{c+i\infty}\frac{\Gamma(s)\Gamma(\gamma-s)}
{\Gamma(\gamma)\Gamma(\beta-\alpha s)}z^{-s}{\rm d}s\\
&=\frac{1}{\Gamma(\gamma)}H_{1,2}^{1,1}\left[z\big\vert_{(0,1),(1-\beta,\alpha)}^{(1,1)},
\right]\tag{6.1}
\end{align*}for $i=\sqrt{-1}, 0<c<\gamma,\alpha>0, \beta>0$. Note that the basic Mittag-Leffler parameter $\alpha$ enters into the H-function as the coefficient of the complex variable $s$. In all the H-function representations of fractional integrals and solutions of fractional differential equations, the fractional index or fractional order $\alpha$, or a function of $\alpha$, enters as the coefficient of the complex variable $s$ in the H-function representation.

\vskip.5cm\noindent{\bf 7.\hskip.3cm Diffusion Equation}

We consider the following diffusion model with fractional-order spatial and temporal derivatives
\begin{equation*}
_0D_t^\beta N(x,t)=\eta\;_xD_\theta^\alpha N(x,t),\tag{7.1}
\end{equation*}
with the initial conditions $_0D_t^{\beta-1}N(x,0)=\sigma(x), 0\leq\beta\leq 1, \lim_{x\to \pm\infty} N(x,t)=0,$ where $\eta$ is a diffusion constant; $\eta,t>0,x\in R; \alpha, \theta, \beta$  are real parameters with the constraints
$$0<\alpha \leq 2, |\theta| \min(\alpha, 2-\alpha),$$
and $\delta(x)$ is the Dirac-delta function. Then for the fundamental solution of (7.1) with initial conditions, there holds the formula
\begin{equation*}
N(x,t)=\frac{t^{\beta-1}}{\alpha|x|}H^{2,1}_{3,3}\left[\frac{|x|}{(\eta t^\beta)^{1/\alpha}}\left|^{(1,1/\alpha), (\beta,\beta/\alpha), (1, \rho)}_{(1,1/\alpha), (1,1), (1,\rho)}\right.\right], \alpha>0\tag{7.2}
\end{equation*}
where $\rho=\frac{\alpha-\theta}{2\alpha}.$
\medskip
\noindent
The following special cases of (1) are of special interest for fractional diffusion models:\\
(i) For $\alpha=\beta$,  the corresponding solution of (7.1), denoted by $N_\alpha^\theta$,  can be expressed in terms of the H-function as given below and can be defined for $x>0$:\par
\medskip
\noindent
Non-diffusion: $0<\alpha=\beta<2; \theta\leq \min \left\{\alpha,2-\alpha\right\},$
\begin{equation*}
N_\alpha^\theta(x)=\frac{t^{\alpha-1}}{\alpha|x|}H^{2,1}_{3,3}
\left[\frac{|x|}{t\eta^{1/\alpha}}\left|^{
(1,1/\alpha),(\alpha,1), (1,\rho)}_{(1,1/\alpha),(1,1), (1,\rho)}\right]\right.,\; \rho=\frac{\alpha-\theta}{2\alpha}.\tag{7.3}
\end{equation*}		
(ii) When $\beta=1,0<\alpha\leq2;\theta\leq \min\left\{\alpha, 2-\alpha\right\}$, then (7.1) reduces to the space-fractional diffusion equation, which is the fundamental solution of the following space-time fractional diffusion model:
\begin{equation*}
\frac{\partial N(x,t)}{\partial t}= \eta\;_xD_\theta^\alpha N(x,t), \eta>0, x\in R,\tag{7.4}
\end{equation*}
with the initial conditions  $N (x,t=0) = \sigma(x), \displaystyle {\lim_{x\to\pm\infty}} N(x,t)=0,$ where $\eta$ is a diffusion constant  and $\sigma(x)$ is the Dirac-delta function. Hence for the solution of (1) there holds the formula
\begin{equation*}
N_\alpha^\theta(x)=\frac{1}{\alpha(\eta t)^{1/\alpha}}\;H^{1,1}_{2,2}\left[\frac{(\eta t)^{1/\alpha}}{|x|}\left|^{(1,1),(\rho, \rho)}_{(\frac{1}{\alpha},\frac{1}{\alpha}),(\rho, \rho)}\right]\right.,\;0<\alpha<1, |\theta|\leq \alpha,\tag{7.5}
\end{equation*}
where $\rho=\frac{\alpha-\theta}{2\alpha}$. The density represented by the above expression is known as $\alpha$-stable L\'{e}vy density. Another form of this density is given by
\begin{equation*}
N_\alpha^\theta (x)=\frac{1}{\alpha(\eta t)^{1/\alpha}}\;H^{1,1}_{2,2}\left[\frac{|x|}{(\eta t)^{1/\alpha}}\left|^{(1-\frac{1}{\alpha},\frac{1}{\alpha}), (1-\rho, \rho)}_{(0,1),(1-\rho, \rho)}\right.\right], 1<\alpha< 2, |\theta|\leq 2-\alpha.\tag{7.6}
\end{equation*}
(iii) Next, if we take $\alpha=2,0<\beta<2;\theta =0$,  then we obtain the time-fractional diffusion,  which is governed by the following time-fractional diffusion model:
\begin{equation*}
\frac{\partial ^\beta N(x,t)}{\partial t^\beta} = \eta \frac{\partial^2}{\partial x^2} N(x,t), \eta>0, x\in R,\; 0<\beta\leq 2,\tag{7.7}
\end{equation*}
with the initial conditions $_0D_t^{\beta-1} N(x,0)= \sigma(x), _0D_t^{\beta-2} N(x,0)=0,\; \mbox{for}\; x \in r, \\
\lim_{x\to\pm\infty} N(x,t)=0$,
where $\eta$  is a diffusion constant  and $\sigma(x)$ is the Dirac-delta function, whose fundamental solution is given by the equation
\begin{equation*}
N(x,t)=\frac{t^{\beta-1}}{2|x|}\;H^{1,0}_{1,1}\left[\frac{|x|}{(\eta t^\beta)^{1/2}}\left|^{(\beta, \beta/2)}_{(1,1)}\right.\right].\tag{7.8}
\end{equation*}
(iv) If we set $\alpha=2, \beta=1$ and $\theta\rightarrow 0$, then for the fundamental solution of the standard diffusion equation
\begin{equation*}
\frac{\partial}{\partial t}N(x,t)=\eta\frac{\partial^2}{\partial x^2}N(x,t),\tag{7.9}
\end{equation*}
with initial condition
\begin{equation*}
N(x,t=0) = \sigma(x), \lim_{x\to \pm\infty} N(x,t)=0,\tag{7.10}
\end{equation*}
there holds  the formula
\begin{equation*}
N(x,t)=\frac{1}{2|x|}H^{1,0}_{1,1}\left[\frac{|x|}{\eta^{1/2}t^{1/2}}
\left|^{(1,1/2)}_{(1,1)}
\right.\right]=(4\pi \eta t)^{-1/2} \exp[-\frac{|x|^2}{4\eta t}],\tag{7.11}
\end{equation*}
which is the classical Gaussian density.\par

\vskip.3cm\noindent\begin{center}{\bf References}\end{center}

\vskip.3cm\noindent[1]~~C. Beck and E.G.D. Cohen (2003): Superstatistics, {\it Physica A}, {\bf 322}, 267-275.

\vskip.2cm\noindent[2]~~A. Carpinteri and F. Mainardi (1997): {\it Fractals and Fractional Calculus in Continuum Mechanics}, Springer-Verlag, Wien.
\vskip.2cm\noindent [3]~~R. Gorenflo, Y. Luchko and F. Mainardi (2000): Wright function as scale invariant solutions of the diffusion-wave equation, {\it Journal of Computational and Applied Mathematics} {\bf 118}, 175-191.

\vskip.2cm\noindent[4]~~H.J. Haubold, A.M. Mathai and R.K. Saxena (2011): Mittag-Leffler functions and their applications, {\it Journal of Applied Mathematics}, ID 298628, 51 pages.
\vskip.2cm\noindent [5]~~H.J. Haubold, A.M. Mathai and R.K. Saxena (2011a): Further solutions of fractional reaction-diffusion equations in terms of the H-function, {\it Journal of Computational and Applied Mathematics}, {\bf 235}, 1311-1316.

\vskip.2cm\noindent [6]~~V.S. Kiriyakova (1994): {\it Generalized Fractional Calculus and Applications}, Wiley, New York.

\vskip.2cm\noindent [7] E. Kr\"atzel (1979): Integral transformations of Bessel type,  In {\it Generalized Functions and Operational Calculus, Proc. Conf. Varna 1975, Bulg. Acad. Sci., Sofia}, 1979, pp. 148-155.

\vskip.2cm\noindent [8]~~A.M. Mathai (1993): {\it A Handbook of Generalized Special Functions for Statistical and Physical Sciences}, Oxford University Press, Oxford.

\vskip.2cm\noindent [9]~~A.M. Mathai (1999): {\it An Introduction to Geometrical Probability: Distributional Aspects with Applications}, Gordon and Breach, Newark.

\vskip.2cm\noindent [10]~~A.M. Mathai (2012): Generalized Kr\"atzel integral and associated statistical densities, {\it International Journal of Mathematical Analysis}, {\bf 6(51)}, 2501-2510.

\vskip.2cm\noindent [11]~~A.M. Mathai (2013): Fractional integral operators in the complex matrix-variate case, {\it Linear Algebra and its Applications}, {\bf 439}, 2001-2013.

\vskip.2cm\noindent [12]~~A.M. Mathai (2014): Fractional integral operators involving many matrix-variables, {\it Linear Algebra and its Applications}, {\bf 446}, 196-215.

\vskip.2cm\noindent [13]~~A.M. Mathai (2014a): {\it An Introduction to Fractional Calculus}, Module 10, CMSS, Peechi, Kerala, India.

\vskip.2cm\noindent [14] ~~A.M. Mathai (2015): Fractional differential operators in the complex matrix-variate case,
{\it Linear Algebra and its Applications}, {\bf 478}, 200-217.

\vskip.2cm\noindent [15]~~A.M. Mathai and H.J. Haubold (1988): {\it Modern Problems in Nuclear and Neutrino Astrophysics}, Akademie-Verlag, Berlin.

\vskip.2cm\noindent [16]~~A.M. Mathai and H.J. Haubold (2007): Pathway model, superstatistics, Tsallis statistics and generalized measure of entropy, {\it Physica A}, {\bf 375}, 110-122.

\vskip.2cm\noindent[17]~~A.M. Mathai and S. B. Provost (2006): Some complex matrix variate statistical distributions in rectangular matrices, {\it Linear Algebra and its Applications}, {\bf 410}, 198-216.

\vskip.2cm\noindent [18]~~A.M. Mathai and S.B. Provost (2006a): On q-logistic and related distributions, {\it IEEE Transactions on Reliability}, {\bf 55)}, 237-244.

\vskip.2cm\noindent [19]~~A.M. Mathai and P.N. Rathie (1975): {\it Basic Concepts in Information Theory and Statistics: Axiomatic Foundations and Applications}, Wiley Halsted, New York and Wiley Eastern, New Delhi.

\vskip.2cm\noindent [20]~~A.M. Mathai, R.K. Saxena and H.J. Haubold (2010): {\it The H-Function: Theory and Applications}, Springer, New York.

\vskip.2cm\noindent [21]~~S.S. Nair (2009): Pathway fractional integration operator, {\it Fractional Calculus \& Applied Analysis}, {\bf 12}, 237-252.

\vskip.2cm\noindent[22]~~ C. Tsallis (2009): {\it Introduction to Nonextensive Statistical Mechanics: Approaching a Complex World}, Springer, New York.

\end{document}